\documentclass[a4paper,11pt]{article}%
\usepackage{amsfonts}
\usepackage{graphicx}
\usepackage{amsmath}
\usepackage{amssymb}%
\usepackage{cite}
\pdfoutput=1
\makeatletter

\@addtoreset{equation}{section}
\makeatother
\newcommand{\beq}{\begin{equation}}
\newcommand{\eeq}{\end{equation}}
\renewcommand{\a}{\alpha}
\renewcommand{\b}{{{\beta}}}
\newcommand{\mi}{\hat{\imath}}
\newcommand{\mj}{\hat{\jmath}}

\newcommand{\be}{\begin{eqnarray}}
\newcommand{\ee}{\end{eqnarray}}

\begin{document}
\baselineskip=18pt
\baselineskip 0.6cm

\begin{titlepage}
\setcounter{page}{0}
\renewcommand{\thefootnote}{\fnsymbol{footnote}}
\begin{flushright}
ARC-18-18
\end{flushright}

\vskip 2cm
\begin{center}
{\Huge \bf  The Quantum Theory \\ of  Chern-Simons Supergravity}
\vskip 1cm
{
\large L.~Andrianopoli$^{a,c,d}$\footnote{laura.andrianopoli@polito.it},
\large B.~L.~Cerchiai$^{a,c,e}$\footnote{bianca.cerchiai@polito.it},
\large P.~A.~Grassi$^{b,c,d}$\footnote{pietro.grassi@uniupo.it}, and
\large M.~Trigiante$^{a,c,d}$\footnote{mario.trigiante@polito.it}
}
\vskip 1cm {
\small
\centerline{$^{(a)}$
\it  DISAT, Politecnico di Torino,}
\centerline{\it C.so Duca degli Abruzzi, 24, 10129 Torino, Italy,}
\medskip
\centerline{$^{(b)}$
\it Dipartimento di Scienze e Innovazione
Tecnologica,} \centerline{\it Universit\`a del Piemonte Orientale,}
\centerline{\it viale T. Michel, 11, 15121 Alessandria, Italy,}
\medskip
\centerline{$^{(c)}$
\it INFN, Sezione di Torino,} \centerline{\it
via P. Giuria 1, 10125 Torino, Italy,}
\medskip
\centerline{$^{(d)}$
\it Arnold-Regge Center,}
\centerline{\it via P. Giuria 1,  10125 Torino, Italy, }
\medskip
\centerline{$^{(e)}$
\it Centro Fermi,}\centerline{\it Museo Storico della Fisica e Centro Studi e Ricerche ''Enrico Fermi'',}
\centerline{\it Piazza del Viminale  1,  00184 Roma, Italy. }
}
\end{center}
\vskip 0.2cm
\centerline{{\bf Abstract}}
\medskip
We consider $AdS_3$ $N$-extended Chern-Simons supergravity (\`a la Achucarro-Townsend) and we study its
gauge symmetries. We promote those gauge symmetries to a BRST symmetry and we perform its quantization
by choosing suitable gauge-fixings. The resulting quantum theories have different features which we discuss in the present
work. In particular, we show that a special choice of the gauge-fixing correctly reproduces the Ansatz by
Alvarez, Valenzuela and Zanelli for the graphene fermion.


\end{titlepage}

\tableofcontents \noindent {}
\newpage
\setcounter{footnote}{0} \newpage\setcounter{footnote}{0}

\section{Introduction}

It is tantalising to relate models which are apparently very different.
Some years ago, Gaiotto and Witten, in \cite{Gaiotto:2008sd}, considered a
Chern-Simons (CS) gauge theory in three dimensions
coupled to $\mathcal{N}=2$ supersymmetric multiplets whose scalar components are coordinates of a hyper-K\"ahler manifold.
The model descends from a four dimensional one in the presence of a defect and the potential is chosen
to enhance the supersymmetry from $\mathcal{N}=2$ to $\mathcal{N}=4$. The conditions under which this is possible are certain
relations between the moment maps associated with  the linear action of the gauge group on the hyper-K\"ahler
manifold which unveil a superalgebra hidden in the model.

They argued that the supersymmetric Wilson loops can be constructed in terms of that super algebra, but they did not elaborate further along these lines. In another paper \cite{Kapustin:2009cd},
Kapustin and Saulina showed that Rozansky-Witten
theory \cite{Rozansky:1996bq} coupled to a Chern-Simons gauge field can be written (up to an exact
BRST term) as a Chern-Simons gauge theory on a supergroup. In particular they showed the following relation
\begin{eqnarray}
\label{Rela_intro}
&&\left. \mbox{Chern-Simons~theory}\right|_{\mathcal{SG}} + s \Big( \mbox{gauge~fixing~of~fermion~gauge~symmetries} \Big) \nonumber \\
&& \hspace{5cm} =  \\
&& \left. \mbox{topologically twisted super-Chern-Simons~theory}\right|_{\mathcal{G}} + {\rm SUSY ~ matter~fields}\,.\nonumber
\end{eqnarray}
where Chern-Simons~theory$|_{\mathcal{SG}}$ is a Chern-Simons theory on the supergroup ${\mathcal{SG}}$
(see also \cite{Mikhaylov:2014aoa} for a complete discussion) plus a gauge-fixing and it can be recast in the
form of a $\mathcal{N}=4$ super-Chern-Simons theory on the
group ${\mathcal{G}}$ (which is the bosonic subgroup of ${\mathcal{SG}}$) coupled to $\mathcal{N}=4$ hyper-multiplets.
The gauginos and the scalar fields of the $\mathcal{N}=4$ Chern-Simons multiplet are not dynamical and their
equations of motion determine them  in terms of the scalars and fermions in the hyper-multiplet sector.
By inserting this solution back in the action, one gets additional non-trivial terms for the potential.

The complete scalar potential in the lower part of the correspondence (\ref{Rela_intro}) is incorporated into the gauge-fixing of the fermionic symmetry inside $\mathcal{SG}$ in the upper part of the same relation. The mapping from the lower to the
upper part has been used in \cite{Drukker:2009hy,Marino:2011nm,Mauri:2018fsf,Bianchi:2018bke}, to compute
supersymmetric Wilson loops in terms of the Chern-Simons theory on the supergroup
instead of the supersymmetric Chern-Simons theory.\par
In the correspondence (\ref{Rela_intro}) the $\mathcal{N}=4$ supersymmetry in the supersymmetric Chern-Simons model originates, via a topological twist, from the BRST invariance of the upper theory in which only the fermionic gauge symmetries are covariantly fixed. It is important to emphasize that, in Chern-Simons theories, this invariance comprises, apart from the ordinary BRST and anti-BRST transformations $s,\,\bar{s}$, also ``vector'' symmetry transformations $s_\mu,\,\bar{s}_\mu$, as found in  \cite{Delduc:1990je} - \cite{Gieres:2000pv}. The \emph{twisted}  $\mathcal{N}=4$ world-volume supersymmetry in the lower-side of the relation (\ref{Rela_intro}) can be understood in terms of the whole set of these invariances, as we shall discuss in the present paper.

One of the purposes of this work is to apply this construction to a specific Chern-Simons theory which describes $\mathcal{N}=2$ supersymmetric ${\rm AdS_3}$ supergravity, as shown by Achucarro and Townsend in \cite{Achucarro:1987vz}.

Indeed, some years ago,  Achucarro and Townsend
observed that, in three dimensions, $N$-extended supergravity in the presence of a negative cosmological constant can
be rewritten as a Chern-Simons theory whose gauge fields take values in the superalgebra $\mathfrak{osp}(p|2) \times \mathfrak{osp}(q|2)$
where $p+q=N$. The bosonic subalgebra is $\mathfrak{so}(p) \otimes \mathfrak{so}(q) \otimes \mathfrak{sp}(2)
\otimes \mathfrak{sp}(2)$ and the
gauge fields associated with  the two $\mathfrak{sp}(2)$'s are given in terms of the vielbein and the spin connection of the three dimensional manifold.
The ``gravitinos'' (the gauge fields associated with the fermionic gauge generators) are in the bifundamental representations of $\mathrm{SO}(p) \times \mathrm{Sp}(2)$ and/or $\mathrm{SO}(q) \times \mathrm{Sp}(2)$ subgroups. The supersymmetry is realized as a gauge symmetry
and therefore the fermionic charges are related to the fermionic (anticommuting)
gauge fields. Being a Chern-Simons theory it does not depend on the 3-dimensional metric on the world-volume. The latter however emerges from the gauge fixing, within BRST-exact terms in the Lagrangian. AdS-supergravity in three dimensions, being a Chern-Simons gauge theory on a supergroup,
is a viable context where to apply the relation proposed by \cite{Gaiotto:2008sd,Kapustin:2009cd}. The structure of Achucarro and Townsend supergravity from a mathematical point of view is reminiscent of the ABJM model \cite{Aharony:2008ug}, which is a difference of two CS actions as well.

 In \cite{Gaiotto:2008sd,Kapustin:2009cd}, the duality is based on a topological twist, where the twist is realized by taking the diagonal subgroup of the product of the original Lorentz group  with an $Sp(2)$ part of the R-symmetry.
On a AdS background, instead, the analogous topological twist  is naturally related to the (non-unique) choice of the Lorentz subalgebra inside the anti-de Sitter isometry $\mathfrak{so}(2,2)\simeq \mathfrak{sp}(2)\times \mathfrak{sp}(2)$. Indeed, in this framework this topological twist corresponds to trading the antisymmetric part of the torsion $\tau$ for a cosmological constant $\Lambda$. More precisely, the choice as Lorentz group of one of the two $Sp(2)$ factors -- which corresponds to the untwisted theory in the flat background -- 
is here associated with a non-vanishing space-time torsion $\tau\neq 0$. On the other hand, choosing instead the diagonal subgroup $SO(1,2)_D \subset SO(2,2)$ as Lorentz group -- which corresponds
to the topologically twisted theory in a flat background --  is here associated with the choice of a torsionless spin-connection in a background with cosmological constant $\Lambda = - \tau^2\neq 0$.

The first step towards constructing the super-Chern-Simons theory coupled to matter fields for 
Achucarro - Townsend supergravity is to perform a suitable
 gauge-fixing of the fermionic part of the super-gauge symmetry. The gauge symmetry at the quantum level is replaced by the BRST symmetry and the
gauge parameters are replaced by the ghost fields which, in the present case in which the gauge fixed symmetries are of fermionic type, are commuting scalar fields.
To complete the gauge-fixing procedure, one needs an auxiliary sector, also known as an anti-ghost sector,
which, in this case, consists of a set of commuting scalar fields and a set of fermionic Nakanishi-Lautrup fields~\cite{Kapustin:2009cd}.
The ghosts and the anti-ghosts are commuting scalar fields belonging to conjugated representations of the bosonic gauge group $\mathcal{G}$. In fact they turn out to span a quaternionic-K\"ahler manifold which carries a tri-holomorphic action of $\mathcal{G}$. Finally, using the work of Kapustin and Saulina \cite{Kapustin:2009cd},
we translate the degrees of freedom of the Chern-Simons supergravity
in terms of the ones of a super-Chern-Simons theory with $\mathcal{N}=4$ extended supersymmetry coupled to matter.
That theory has a scalar potential, which can be related to the gauge-fixing of the Chern-Simons on the supergroup.\par

Of course, there are several gauge-fixing choices and we will explore them, pointing out the relevant
features of the corresponding quantized theories. In the BRST formalism the gauge-fixing is chosen by adding to the action the BRST variation of the gauge-fixing fermion $\Psi$.
The latter has to carry negative ghost number, it should be Lorentz invariant and, since at this stage we are gauge-fixing only the fermionic
gauge symmetries, it has to be gauge invariant under the bosonic symmetries.

In particular, we point out that, among the possible gauge-fixing choices, there is an unconventional one, whose degrees of freedom correspond to a propagating massive Dirac spinor, which reproduces the field content of the model described by Alvarez, Valenzuela and Zanelli  \cite{Alvarez:2011gd,Guevara:2016rbl}, to be referred to in the sequel as AVZ model. The latter is based on an $N=2$ supergroup and provides a phenomenological description of graphene. In that case, a three dimensional Chern-Simons theory with ${\mathrm{OSp}}(2|2)$ gauge group, and the fermionic 1-forms $\psi_{I}^\alpha$, $I=1,2$, in the bifundamental of the ${ \mathrm{Sp}}(2)\times {\mathrm{SO}}(2)$ group are written in terms of spin-$1/2$ fields $\chi^\alpha_I$ through of the Ansatz:
\begin{equation}
\psi_{I}^\alpha = i\, e^i\,(\gamma_i)^\alpha{}_\beta\,\chi^\beta_I\,,\label{ZansaZ}
\end{equation}
where $e^i$, $i=0,1,2$, are the vielbein 1-forms of the three dimensional spacetime and $\gamma^i$ are the corresponding gamma matrices. Since $e^i$ and $\chi^\beta_I$ only enter the action through the above Ansatz, the theory is invariant under the local rescaling symmetry \cite{Alvarez:2011gd}:
 $e^i\rightarrow \lambda(x) e^i\,,\,\,\,\chi^\beta_I\rightarrow \frac{1}{\lambda(x)}\,\chi^\beta_I$, $\lambda(x)$ being a real function.
The ${ \mathrm{Sp}}(2)$-connection is identified with the Lorentz one $\omega^{ab}$, and a space-time torsion $T^i=\mathcal{D}e^i$ is allowed for. By suitably fixing the local rescaling symmetry of $e^i$, $T^i$ can be made constant of the form:
\begin{equation}
T^i=\mathcal{D}e^i\equiv de^i+\omega^i{}_j\wedge e^j=\tau\,\epsilon^i{}_{jk}\,e^j\wedge e^k\,,
\end{equation}
$\tau $ being a dimensionful constant.\par
The Ansatz (\ref{ZansaZ}) amounts to setting the spin-$3/2$ component of the gravitino fields to zero, keeping however a non-zero spin-$1/2$ component. As a consequence of this choice, the original Chern-Simons theory yields an effective model describing a propagating massive spin-$1/2$ Dirac field $\chi^\alpha=\chi^\alpha_1+i\,\chi^\alpha_2$, whose mass is related to the spacetime torsion $\tau$. This model is suited to describe graphene in the presence of space-time curvature and torsion.\par In \cite{Andrianopoli:2018ymh} the model of \cite{Alvarez:2011gd} is embedded in supergravity. First of all it is embedded in $AdS_3$ supergravity by identifying its gauge supergroup with the $\mathrm{OSp}(2|2)$-factor of the super-$AdS_3$ symmetry ${ \mathrm{OSp}}(2|2)\times { \mathrm{SO}}(2,1)$. The $D=3$ supergravity is then characterized as the boundary theory of an $AdS_4$ supergravity with ${N}=2$ supersymmetry \cite{Andrianopoli:2014aqa}. In this holographic correspondence, an appropriate parametrization of the $AdS_4$ space is chosen, which corresponds to an $AdS_3$-slicing of the same space. Furthermore, by choosing suitable boundary conditions for the four-dimensional fields, the model of \cite{Alvarez:2011gd} is retrieved at the $AdS_3$ boundary. In this picture the spin-$1/2$ field $\chi^\alpha$, which ought to describe the collective electron modes in the graphene, originates from the \emph{radial} component of the $D=4$ gravitino field (i.e. the component of the gravitino 1-form along the direction perpendicular to the boundary), and the torsion parameter $\tau$, which behaves as a mass term for the spinor $\chi$,  is naturally related to the curvature of the $AdS_3$ spacetime. However, while in the AVZ model of \cite{Alvarez:2011gd,Guevara:2016rbl} the presence of a cosmological constant, with the corresponding enhancement of the gauge symmetry to $\mathrm{OSp}(2|2)\times \mathrm{SO}(1,2)$, is optional, this is not the case if one aims to identify the CS model with D=3 supergravity, since the Achucarro-Townsend map \cite{Achucarro:1987vz}, on which the identification in \cite{Andrianopoli:2018ymh}  is based, requires a non-vanishing cosmological constant, which induces a non-vanishing mass term for the Dirac spinor $\chi$.
\par
In both the constructions in \cite{Alvarez:2011gd} and  \cite{Andrianopoli:2018ymh}, the condition (\ref{ZansaZ}) is put by hand. An important goal of the present paper is to retrieve it dynamically within a covariant BRST-quantized setting.
As said above, by a  Landau-type gauge-fixing of the (gauge) supersymmetry, we are able to see that there is one massless Dirac spinor propagating. However, to
compare it with the Ansatz (\ref{ZansaZ}) for the case of a massive spinor, as it is the case in \cite{Andrianopoli:2018ymh}, we have to modify the gauge-fixing term by adding
a first order differential and a vorticity term to the action. By a simple analysis of the quadratic part of the action,
it is then easy to show that the Dirac spinor has a non-vanishing mass related to the cosmological constant.

Another outcome of our analysis is the study of the symmetry properties of the quantum model on the lower-side of the relation in (\ref{Rela_intro}). In particular we find, at least for the conventional Landau gauge-fixing, that the quantum CS-theory exhibits a rigid $\mathcal{N}=4$ world-volume supersymmetry on $AdS_3$, which comprises, besides the BRST symmetry, also an emerging ``vector''-supersymmetry \cite{Delduc:1990je} - \cite{Gieres:2000pv}. We shall expand on this particular issue in a forthcoming paper \cite{long_paper}.

The paper is organized as follows:
In section 2 we define the classical symmetries of the $D=3$ Chern-Simons supergravity and its BRST symmetries. In particular
in subsect. 2.1 we recall the basic facts about three dimensional AdS $N$-extended supergravity. In subsect. 2.2 we review its interpretation as describing the graphene in the AVZ Ansatz. In subsect. 2.3, we present the BRST formulation. In subsect.s 2.4 and 2.5, the secondary BRST symmetry $\bar{s}$ is
discussed as well as the ``vector'' BRST symmetry transformations $s_\mu,\,\bar{s}_\mu$.\\
In sect. 3, we finally quantize the model by considering different types of gauge-fixing:\\
1) a conventional gauge-fixing leading to a massless Dirac spinor (Landau gauge-fixing) and yielding a dual ${\cal N}=4$
supersymmetric model,\\
2) Landau gauge-fixing with additional non-linear terms to reproduce the scalar potential in the dual theory,\\
3) an additional term (Nakanishi-Lautrup term) to allow for mass deformations of the model,\\
4) an $s \bar s$ gauge-fixing based on the presence of a secondary BRST symmetry $\bar s$ and, finally,\\
5) an unconventional gauge-fixing, which reproduces the AVZ Ansatz with a non-vanishing mass.\\

We conclude with the summary and with the future perspectives.

\section{D=3 $N$-Extended Chern-Simons Supergravity}

\subsection{Basic Facts}

We first recall some  basic facts about $D=3$ supergravity with negative cosmological constant. As is well-known from the work of Achucarro and
Townsend \cite{Achucarro:1987vz}, $N$-extended D=3 supergravity on $AdS_3$
can be formulated in terms of a Chern-Simons gauge theory. More precisely, in the absence of non-trivial boundary conditions, it can be rephrased as the difference of two Chern-Simons Lagrangians, associated, respectively, with the supergroups $G_+=\mathrm{OSp}(p|2)$ and  $G_-=\mathrm{OSp}(q|2)$, with $p + q =N$  (and bosonic parts $\mathrm{SO}(p) \times \mathrm{Sp}(2)_{(+)}$ and  $\mathrm{SO}(q) \times \mathrm{Sp}(2)_{(-)}$, respectively):
\begin{equation}
\mathcal{L}_{SUGRA}=\mathcal{L}_{CS}^{(G_+)}-\mathcal{L}_{CS}^{(G_-)}\,.\label{atcs}
\end{equation}
The gauge connections of the two CS theories are
\begin{eqnarray}
\mathcal{A}_{(+)} &=& \frac 12 \omega_{(+)}^{\imath\jmath}\mathbb{J}_{\imath\jmath} +A_{(+)}^{IJ}\mathbb{T}_{IJ}+ \bar{\mathbb{Q}}^{\alpha|I}\psi^{(+)}_{\alpha | I}\\
\mathcal{A}_{(-)} &=& \frac 12 \omega_{(-)}^{\mi\mj}\mathbb{J}_{\mi\mj} +A_{(-)}^{\dot I\dot J}\mathbb{T}_{\dot I\dot J}+ \bar{\mathbb{Q}}^{\dot\alpha|\dot I}\psi^{(-)}_{\dot\alpha | \dot I}\,.
\end{eqnarray}
Here $\mathbb{J}_{\imath\jmath}$ ($\imath,\jmath=0,1,2$),  $\mathbb{J}_{\mi\mj}$ ($\mi,\mj=0,1,2$) are the generators of $\mathfrak{so}(1,2)_{(+)} \sim \mathfrak{sp}(2)_{(+)}$ and $\mathfrak{so}(1,2)_{(-)} \sim \mathfrak{sp}(2)_{(-)}$, respectively,
$\mathbb{T}_{IJ}$ ($I,J=1,\cdots p$), $\mathbb{T}_{\dot I\dot J}$ ($\dot I,\dot J=1,\cdots q$) are the generators of $\mathrm{SO}(p)$ and $\mathrm{SO}(q)$, respectively, while ${\mathbb{Q}}^{\alpha|I}$, ${\mathbb{Q}}^{\dot\alpha|\dot I}$ ($\alpha=1,2\in \mathrm{Sp}(2)_{(+)}$, $\dot\alpha=1,2\in \mathrm{Sp}(2)_{(-)}$) are the (Majorana) fermionic generators of the two supergroups.\footnote{We denote with a bar the adjoint fermion: $\bar{\mathbb{Q}}\equiv  {\mathbb{Q}}^t \ C$, $C$ being the charge conjugation matrix.}
Finally, $\omega_{(\pm)}$, $A_{(\pm)}$, $\psi_{(\pm)}$ denote the corresponding gauge connections, the last being associated with Majorana spinor 1-forms.

The relation between the topological CS theory and D=3 $N$-extended supergravity (which does not have local propagating degrees of freedom) with $AdS_3$ radius $\ell$, is found by introducing the fields:
\begin{eqnarray}
\omega^{ij}&=&\frac{1}{2}\left(\omega_{(+)}^{\imath\jmath}+\omega_{(-)}^{\mi\mj}\right)\label{omegadiag}
\\
E^k &=&\frac{\ell}{4}\left(\omega_{(+){\imath\jmath}}-\omega_{(-)\mi\mj}\right)\epsilon^{ijk}\label{Ediag}
\end{eqnarray}
where $\omega^{ij}$ is identified with the (torsionless) spin connection of the Lorentz algebra: $$\mathfrak{so}(1,2)_D\subset \mathfrak{so}(2,2) = \mathfrak{so}(1,2)_{(+)}\times \mathfrak{so}(1,2)_{(-)}$$ and $E^i$ as the bosonic components of the supervielbein of the $N$-extended superspace. Note that, in eqs.~(\ref{omegadiag}) and (\ref{Ediag}), the identification of the indices $\imath,\jmath,...$ and $\mi,\mj,...$ with the anholonomic Lorentz indices $i,j,...$ is understood in the definition of the spin connection  and dreibein of $D=3$  supergravity. This corresponds to the fact that the supergravity Lagrangian exhibits manifest invariance  with respect to the diagonal Lorentz group $\mathrm{SO}(1,2)_D\subset \mathrm{SO}(2,2)$.

\vskip 5mm
Recently, some of us reconsidered, in \cite{Andrianopoli:2018ymh}, the Achucarro-Townsend theory \cite{Achucarro:1987vz} and the  correspondence (\ref{atcs}) for the special case $N=p=2$, $q=0$. In particular, the field equations of the $N=2$ $AdS_3$ supergravity  were found as asymptotic boundary conditions on the supergravity field-strengths of $N=2$ $AdS_4$ pure supergravity,
 along the lines  discussed in  \cite{Andrianopoli:2014aqa}.
 In that framework, $\mathrm{OSp}(2|4)$-invariant Neumann conditions are recovered on the boundary as consistency conditions for supersymmetry of the full action. An $N=p=2$ $AdS_3$ description of those Neumann conditions  was found, in \cite{Andrianopoli:2018ymh}, for an asymptotic boundary located at $r\to\infty$ as a particular asymptotic limit, inspired by the so-called ``ultraspinning limit''~\cite{Caldarelli:2008pz}
in the Fefferman-Graham parametrization of the $D=4$ superfields.

 The resulting $N=p=2$, $D=3$  supergravity Lagrangian reads:
 \begin{equation}\label{lag3D}
 \mathcal{L}_{(3)}= \left(\mathcal R^{ij} -\frac{1}{3\,\ell^2}  E^i E^j  -\frac{1}{2\,\ell} \, \bar\psi_I\gamma^{ij}\psi_I\right) E^k \epsilon_{ijk} - \frac{1}{2\,\ell} AdA +2 \bar\psi_I\left(\mathcal{D} \psi_I - \frac{1}{2\ell}\,\epsilon_{IJ}\,A \,  {\psi}_{J}\right)\,,
\end{equation}
where $i,j,...=0,1,2 \in \mathrm{SO}(1,2)_D\subset \mathrm{SO}(2,2)$, $I,J=1,2 \in \mathrm{SO}(2)$. (The $\mathrm{SO}(2)$ repeated indices are meant to be summed over, independently of their position). Its equations of motion are easily written, using (\ref{omegadiag}), (\ref{Ediag}), as the $\mathrm{OSp}(2|2)_{(+)}\times \mathrm{SO}(1,2)_{(-)}$ Maurer-Cartan equations:
\begin{eqnarray}
\mathcal{R}^{\imath\jmath}_{(+)}&=&\, \frac{\rm i}{\ell}\,\bar{\psi}_I\wedge\gamma_{k}\psi_I\,\epsilon^{\imath\jmath k}\,, \label{asymptoticRip}
\nonumber \\
{\mathcal{D}_{(+ )}}\psi_I&=&\frac{1}{2\ell}\,A \wedge  \,\epsilon_{I J}  {\psi}_{J}\,,\label{asymptoticpsi} \nonumber\\
dA &=& \epsilon_{I J}\bar{\psi}_{I}\wedge\psi_{J}\, , \label{asymptoticA} \nonumber \\
\mathcal{R}^{\mi\mj}_{(-)}&=&0 \,,\label{asymptoticRim}
\end{eqnarray}
where:
\begin{eqnarray}
 R^{\imath\jmath}_{(+)} \equiv d\omega^{\imath\jmath}_{(+)} + \omega^{\imath k}_{(+)}\wedge\omega_{(+)k}{}^{\jmath}\,, &&
 R^{\mi\mj}_{(-)} \equiv d\omega^{\mi\mj}_{(-)} + \omega^{\mi \hat k}_{(-)}\wedge\omega_{(-)\hat k}{}^{\mj}\,, ~~~~
A^{IJ}_{(+)} = \epsilon^{IJ}A\,.
\end{eqnarray}

\subsection{A Model for Graphene from D=3 Supergravity}
\label{girella}

The  aim of the analysis in \cite{Andrianopoli:2018ymh} was to make contact with the results of \cite{Alvarez:2011gd}, where the $D=3$ Chern-Simons theory of the supergroup $\mathrm{OSp}(2|2)_{(+)}$ was considered, assuming however a peculiar Ansatz for the odd component of the gauge connection 1-form:
\begin{equation}\label{zanelli}
  \psi^{\alpha}_I={\rm i}\left(\gamma_i\right)^{\alpha}_{~\beta} \chi^{\beta}_I e^i\,.
\end{equation}
Here, $e^i, \gamma^i$ are, respectively, the dreibein and  a set of gamma matrices on the $D=3$ world-volume where the Chern-Simons theory is defined. With the assumption (\ref{zanelli}), the CS Lagrangian turns out to describe the local dynamics of the spin-1/2 field $\chi\equiv  \chi_{I=1}+{\rm i}\chi_{I=2}$. More precisely, $\chi$ is a Dirac spinor satisfying in general the massive Dirac equation, with mass given in terms of the contorsion $\tau=\frac 1{6}\epsilon^{ijk}(De_{[i})_{jk]}$. For non-zero $\tau$, as discussed in \cite{Alvarez:2011gd}, the contorsion can be set to zero by a redefinition of the spin connection, and with that choice  the background space-time turns out to be $AdS_3$, with cosmological constant $-\tau^2$, and the world-volume symmetry  is enhanced to $\mathrm{SO}(2,2)'$.\footnote{Here and in the following, we shall distinguish by a prime quantities referred to the world-volume from the analogous quantities on the target space.}

In \cite{Andrianopoli:2018ymh} it was shown that, in the  case of contorsion $\tau = -\frac 1\ell$, the model of \cite{Alvarez:2011gd} can be recovered at the asymptotic boundary of $AdS_4$, $N=2$ supergravity. It corresponds to imposing in a non-trivial way the condition that in $D=4$ supergravity projects out the spin-1/2 part of the gravitino field:
\begin{equation}\label{psir}
  \Gamma^{\hat\mu}\Psi_{\hat{\mu}I}=0 \,\Rightarrow \gamma^\mu\psi_{\mu I} = -\gamma^r\psi_{rI}= 3{\rm i}\chi_I\neq 0
\end{equation}
where $\Gamma,\Psi$ denote $D=4$ gamma-matrices and gravitino, respectively, $\hat\mu=(\mu,r)=0,1,2,3$ being holonomic world indices.

The Ansatz (\ref{zanelli}) of \cite{Alvarez:2011gd}, in light of its relation with supergravity in $D=3$ \cite{Achucarro:1987vz} and in $D=4$ \cite{Andrianopoli:2018ymh}, is remarkable in several respects:
\begin{itemize}
\item It introduces in the topological Chern-Simons Lagrangian a dependence on the space-time background and a local dynamics for the spinor $\chi$.
\item It implies that the radial component of the $D=4$ gravitino is not suppresed in the asymptotic limit.
\item It also implies a non-trivial relation between the world-volume dreibein $e^i$ and the bosonic part of the super-dreibein $E^i$, which is discussed in detail in \cite{Andrianopoli:2018ymh}.
\end{itemize}
When writing the Ansatz (\ref{zanelli}), a clear distinction has to be made between the target space quantities, which are connections of the superalgebra, and the world-volume quantities $e^i$, $\chi$, and the world-volume Lorentz connection. It is appropriate, therefore, to introduce a different notation for the spinorial index on the world-volume, denoting it by ``primed'' greek letters. Accordingly, eq. (\ref{zanelli}) will then be written, in the following, as:
    \begin{equation}\label{zanelliwv}
  \psi^{\alpha}_I={\rm i}\left(\gamma_i\right)^{\alpha}_{~\beta'} \chi^{\beta'}_I e^i\,.
\end{equation}
In this expression it is manifest that $\left(\gamma_i\right)^{\alpha}_{~\beta'}$ is an intertwiner between the spinor representation of the target space, labeled by $\alpha$, and the one on the world-volume, labeled by $\beta'$. An identification between the Lorentz groups on the target space and on the world-volume is implicit. In flat space, this identification would be unambiguous.  However, in the Achucarro-Townsend model a negative cosmological constant is present on the target space, and this naturally induces the same anti-de Sitter geometry also on the world-volume. This allows multiple choices for the identification of the two Lorentz groups inside the two $SO(2,2)$ $AdS_3$ symmetry groups, which, due to the non trivial relationship between the $E^i$ and $e^i$, are distinct.

We can identify the Lorentz group, both in the target space and in the world-volume, with the diagonal $SL(2)_D\subset SO(2,2)$, which is associated with a Riemannian  spin-connection.
Alternatively, we can identify the common Lorentz group with one of the two   $SL(2)_\pm$ factors. The corresponding spin-connection is torsionful; this is the choice made in \cite{Andrianopoli:2018ymh}. In this latter case, the $SL(2)\subset SO(2,2)$ factor  which is not identified with the Lorentz group can be interpreted as an internal symmetry, associated with new spinorial indices: $\dot\alpha$ in the target space and $\dot\alpha'$ on the world-volume.
This observation will be relevant for the discussion in the next sections.

In the forthcoming section, we will show  that the condition (\ref{zanelli}), with all its peculiar properties discussed above, can be naturally reproduced as a (non-standard) gauge-fixing of the gauged supersymmetry of the Chern-Simons theory, or equivalently, in light of the correspondence in (\ref{atcs}), of $D=3$ supergravity. In the following we are going to reformulate the theory in a BRST covariant framework in order to set up the gauge-fixing properly.

\subsection{BRST Formulation of $N=2$ $AdS_3$ Supergravity}

In the following we will find useful to keep manifest only the Lorentz subalgebra $\mathfrak{sp}(2)_D\sim \mathfrak{so}(1,2)_D\subset \mathfrak{so}(1,2)_{(+)}\times \mathfrak{so}(1,2)_{(-)} $, as discussed above. To this aim we introduce the $\mathfrak{so}(1,2)_D$-covariant notations  for the $\mathfrak{so}(1,2)_{(+)}\times \mathfrak{so}(1,2)_{(-)}$ spin connections:
\begin{equation}
\omega^{\alpha\beta}_{\pm}\equiv \frac 12 \gamma_{ij}^{\alpha\beta}\left(\omega^{ij}\pm \frac 1\ell E_k\epsilon^{ijk}\right)
\end{equation}
and for the corresponding field strengths  $\mathcal{R}^{\alpha\beta}_{\pm}= d \omega_\pm^{\a\b}  - \frac12 \omega^{\a\gamma}_\pm
\epsilon_{\gamma\delta} \wedge \omega^{\delta\b}_{\pm}$. From now on the identification of the spinor indices $\alpha=\dot\alpha=1,2 \in \mathrm{Sp}(2)_D$ is understood. In addition, we rescale
the fields as follows: $\psi \rightarrow \sqrt{\frac\ell 2} \psi$ and $A \rightarrow \ell A$. 

The equations of motion then read:
\begin{eqnarray}
&&{\cal R}^{\a\b}_+ = -\,i\,\, \psi^{\a I} \wedge  \psi^\b_I \,, \qquad
({\cal D} \psi)^\a_I = \frac12 A \wedge \epsilon_{I J} \psi^\a_J\,, \qquad
d A =-\,i\,\frac 12 \epsilon^{IJ} \epsilon_{\a\b} \psi^\a_I \wedge  \psi^\b_J \,,\nonumber\\
&&{\cal R}^{\a\b}_- =0\,,
\end{eqnarray}
where the Lorentz covariant derivative is ${\cal D} \psi^\a_I = d \psi^\a_I + \frac12 (\omega_+)^\a_{~\b} \psi^\b_I$. \footnote{
The $\mathrm{SO}(2)$ indices $I,J,...$ are lowered and raised with a Kronecker delta, which is generally omitted. We always assume that repeated indices are summed over, independently of their position.
As far as doublet $\mathrm{SL}(2,\mathbb{C})$-indices are concerned, they are raised and lowered by the $\epsilon$ symbol, using the ``NE - SW'' convention:
$\xi^\a= \xi_\b \epsilon^{\b\a}\,,\xi_\a= \epsilon_{\a\b} \xi^\b$.
As for the signature of space-time metric, we use mostly minus convention,  as in \cite{Andrianopoli:2018ymh}. Finally for the conjugation of Grassmann numbers we use the convention: $(\xi \lambda)^*=\lambda^*\,\xi^*$.}

The supersymmetry transformations can be cast in the following way
\begin{eqnarray}
\label{EQCICCIO}
&&\delta \omega^{\a\b}_- =0\,, ~~~~ \nonumber \\
&&\delta \omega^{\a\b}_+ =-\,2\,i\, \epsilon^{(\a I} \psi^{\b)}_I  \,, ~~~~\nonumber \\
&&\delta A = -\,i\, \epsilon^{IJ} \epsilon_{\a\b} \epsilon^\a_I \psi^\b_J\,, ~~~~\nonumber \\
&&{\delta} \psi^\a_I = {\cal D} \epsilon^\a_I - \frac12 A \epsilon_{IJ} \epsilon^\a_J \equiv \nabla  \epsilon^\a_I \,.
\end{eqnarray}
The supersymmetry parameter $\epsilon^\a_I$ is a local fermionic real parameter.
Although we are interested in the quantization of the full gauge symmetry, which requires the gauge-fixing of the full superalgebra, in the present paper we only focus on the generators associated with the supercharges. The corresponding ghosts will be interpreted as scalar fields in the dual picture. The
gauge-fixing of the rest of the gauge symmetry is performed along the conventional procedure.

If we promote the local supersymmetry parameter $\epsilon^\a_I$
to a quantum field, it becomes a ghost field that we denote by $\phi^\a_I$. Note
that, since $\phi^\a_I$ is the ghost field of the supersymmetry, it has opposite statistics and, therefore,
it is a commuting scalar field. On the other hand, it carries a positive ghost charge with respect to
a corresponding $U(1)$ group. Then, $\phi^\a_I$ is intrinsically complex, but appears only holomorphically in the
action.

We can translate (\ref{EQCICCIO}) into BRST transformation rules:
\begin{eqnarray}
\label{EQCA}
&&s\, \omega^{\a\b}_- =0\,, ~~~~\nonumber \\
&&s\,  \omega^{\a\b}_+ =-2 \,i\, \phi^{(\a I} \psi^{\b)}_I  \,, ~~~~\nonumber \\
&&s\,  A = -\,i\, \epsilon^{IJ} \epsilon_{\a\b} \phi^\a_I \psi^\b_J\,, ~~~~\nonumber \\
&&s\, \psi^\a_I = {\cal D} \phi^\a_I - \frac12 A \epsilon_{IJ} \phi^\a_J \equiv \nabla  \phi^\a_I \,, \nonumber \\
&&s\, \phi^\a_I =0\,.
\end{eqnarray}
where we set the  BRST transform of the ghost field $\phi^\a_I$ to zero, since we are only
dealing with fermionic gauge symmetries.

Let us also check the nilpotency of the BRST transformations.
By introducing the two composite fields (one for each bosonic generator of the supergroup, namely $\mathrm{Sp}(2)$ and $\mathrm{SO}(2)$)
\begin{eqnarray}
\label{EQCC}
\mu^{(\a\b)}_+ = -\,\phi^{\a }_I \phi^{\b}_I\,, ~~~~~~
\mu_{+ [IJ]} = - \epsilon_{\a\b} \phi^{\a}_I \phi^{\b}_J= \epsilon_{IJ}\,\mu_+\,,
\end{eqnarray}
(the notation is adopted from \cite{Kapustin:2009cd} where $\mu_+^{(\a\b)}$ and $\mu_+$
denote the holomorphic moment
maps of the action of the gauge group $\mathrm{Sp}(2) \times \mathrm{SO}(2)$ on the vector space of the ghost fields
$\phi^\a_I$)
we
find
\begin{eqnarray}
\label{EQCD}
&&s^2 \, \omega^{\a\b}_- =0\,, \nonumber \\
&&s^2 \,  \omega^{\a\b}_+ = \,i\,\nabla \mu^{(\a\b)}_+\,, \nonumber \\
&&s^2 \, A  = \,i\, \nabla \mu_+\,, \nonumber \\
&&s^2 \, \psi^\a_I = \frac i2 \mu^+_{IJ} \psi^\a_J +\frac i2\mu^{+|(\a\b)}\epsilon_{\b\gamma}\psi^\gamma_I\,,
\nonumber \\
&&s^2 \, \phi^\a_I =0\,.
\end{eqnarray}
Note that the above BRST transformations are not nilpotent (except those on $\omega_-$ and on
$\phi^\a_I$), but they yield bosonic gauge transformations of the Lie algebra
$\mathfrak{osp}(2|2)_{(+)}$
with local parameters $-i\,\mu^{(\a\b)}_+$ and $-i\,\mu_+$.
A nilpotent BRST symmetry is attainable by adding the ghosts $c^{(\a\b)}$ and
$c$ of the bosonic symmetry $\mathrm{Sp}(2) \times \mathrm{SO}(2)$.
That follows the conventional procedure and we refer to the vast literature on the subject, see for instance
\cite{Kapustin:2009cd}. For the purpose
of the present paper, we do not need to describe this sector and therefore we omit it.

To set up the gauge-fixing, one needs also some BRST doublets\footnote{A BRST doublet is cohomologically trivial and this implies that all observables are independent of it.} which transform in the
conjugate representation with respect to  $\psi^\a_I$ and $\phi^\a_I$. To this aim we introduce the set
$(\bar \phi^I_\a, \bar \eta^I_\a)$\footnote{Beware: the bar over the fields in the BRST-exact sector denotes the anti-ghost sector.}
with the BRST tranformations:
\begin{eqnarray}
\label{EQCE}
s\, \bar \phi^I_\a = \bar \eta^I_\a\,, ~~~~~~
s\, \bar \eta^I_\a = \frac i2 ( \mu^+_{IJ} \overline\phi_{\a J} +\mu^{+|(\delta\b)}\epsilon_{\a\delta}\overline\phi_{\b I} )\,.
\end{eqnarray}
One can verify that, acting twice with the BRST differential $s$ on the latter fields, one has
again nilpotency up to gauge transformations (as in eq. (\ref{EQCD})).
With the anti-ghost fields $\bar\phi^I_\a$, we can define the moment maps related to the
K\"ahler structure
$K = d\phi^\a_I \wedge d\bar\phi^I_\a$ as
follows:
\begin{eqnarray}
\label{GFC}
\mu_{3}^{\a\b} =- \bar \phi^{(\a}_I \phi^{\b)}_I\,, ~~~~~~
\mu_{3|IJ} = - \phi^{\a}_{[I} \bar \phi^{\b}_{J]}\epsilon_{\a\b}\,.
\end{eqnarray}
where, again, the notation is  adopted form \cite{Kapustin:2009cd}.
In addition to the holomorphic moment maps given in (\ref{EQCC}), we can also introduce the anti-holomorphic moment maps:
\begin{eqnarray}
\label{EQCCanti}
\mu^{(\a\b)}_- = -\,\bar\phi^{\a }_I\bar \phi^{\b}_I\,, ~~~~~~
\mu_{-|[IJ]} = - \epsilon_{\a\b} \bar\phi^{\a}_I \bar\phi^{\b}_J= \epsilon_{IJ}\,\mu_-\, .
\end{eqnarray}
There is a hyper-K\"ahler structure underlying the above relations (\ref{EQCC}), (\ref{GFC}), (\ref{EQCCanti}). Indeed, the scalar bosonic ghost fields $\phi$ and $\bar\phi$ introduced for the gauge-fixing have a natural interpretation
as coordinates on a hyper-K\"alher manifold, as emphasized in \cite{Kapustin:2009cd}, where the $SU(2)$ symmetry associated with the hyper-K\"ahler structure can be made manifest by arranging $\phi^\alpha_I$ and $\bar \phi^I_\alpha$ in the following doublet $\Phi^{\alpha | A}_I$, $A=1,2$:
\begin{equation}
\Phi^{\alpha | 1}_I= \phi^\alpha_I\,,\qquad \Phi^{\alpha | 2}_I= -\epsilon^{\a\b}\bar \phi_{\b I}\label{Phi}\,.
\end{equation}
Here $A$ labels the eigenvectors of the $U(1)$ generator having eigenvalues $\pm i$ on the ghost and anti-ghost field, respectively. The hyper-K\"ahler structure is described by the following triplet of  closed 3-forms $\Omega^{AB}=\Omega^{BA}$:
\begin{equation}
\Omega^{AB}=\epsilon_{\alpha\beta}\, d\Phi_I^{\alpha A}\wedge d\Phi_I^{\beta B}\,.
\end{equation}
We shall also denote by $\Omega=\Omega^{11}=\epsilon_{\alpha\beta}\, d\phi_I^{\alpha}\wedge d\phi_I^{\beta}$ and $\overline{\Omega}=\Omega^{22}=\epsilon_{\alpha\beta}\, d\bar{\phi}_I^{\alpha}\wedge d\bar{\phi}_I^{\beta}$ the holomorphic and anti-holomorphic structures, respectively. The K\"ahler form $K$ introduced earlier, on the other hand, coincides with the remaining component of $\Omega^{AB}$: $K=\Omega^{12}$.
In terms of $\Omega^{AB}$, the triholomorphic moment maps, associated with the generators of $\mathrm{Sp}(2)\times \mathrm{SO}(2)$ symmetry group, are defined as follows:
\begin{equation}
\iota_{V^{(\alpha\beta)}}\Omega^{AB}=-d\mu^{AB|(\alpha\beta)}\,,\,\,\,\iota_{V_{IJ}}\Omega^{AB}=-d\mu^{AB}_{IJ}\,,
\end{equation}
$V^{(\alpha\beta)},\,V_{IJ}$ being the Killing vectors generating $\mathrm{Sp}(2)$ and $\mathrm{SO}(2)$, respectively.\footnote{In our notations:
\begin{equation*}
\delta_{\mathrm{Sp}(2)}\Phi^{\gamma\,A}_I=\frac{1}{2}\,\lambda_{(\alpha\beta)}(V^{(\alpha\beta)})^{\gamma\,A}_I=\frac{1}{2}\,\lambda_{(\alpha\beta)}\epsilon^{\gamma(\alpha}\Phi^{\beta)\,A}_I\,\,,\,\,\,\,
\delta_{\mathrm{SO}(2)} \Phi^{\alpha\,A}_I=\frac{1}{2}\,\lambda^{KL}(V_{KL})^{\alpha\,A}_I=-\frac{1}{2}\,\lambda^{KL}\delta_{I[K}\Phi^{\alpha\,A}_{L]}\,.
\end{equation*}}
The explicit form of the triholomorphic moment maps is readily computed to be:
\begin{eqnarray}
\label{EQcov}
\mu^{AB | (\a\b)} = - \Phi^{(\a | A }_I \Phi^{\b ) B}_I\,, ~~~~~~
\mu^{AB}_{IJ} = - \epsilon_{\a\b} \Phi^{\a |A}_{[I}  \Phi^{\b | B}_{J]}=\epsilon_{IJ}\mu^{AB}\,.
\end{eqnarray}
In particular, with reference to the above definition, we have the following identifications:
\begin{eqnarray}
&\mu^{11 | (\a\b)}= \mu^{(\a\b)}_+\,; \quad \mu^{22 | (\a\b)}=\epsilon^{\a\gamma}\epsilon^{\b\delta} \mu_{- (\gamma\delta)}\,; \quad \mu^{12 | (\a\b)}= \mu^{(\a\b)}_3;&\nonumber\\
&\mu^{11}= \mu_+\,; \quad \mu^{22 }= \mu_-\,; \quad \mu^{12  }= \mu_3\, .&
\end{eqnarray}
In addition, the moment maps satisfy
the condition: \footnote{The symmetric couple of indices $\a\b$ is lowered by the Cartan-Killing metric $k_{(\a\b) (\gamma\delta)}= \epsilon_{\a(\gamma}\epsilon_{\delta)\b}$.}
\begin{eqnarray}
\label{EQCCantiA}
\mu^{(\a\b)}_+ \mu_{+| (\a\b)} + 2 \mu_+^2  =0\,,
\end{eqnarray}
that is crucial for the closure of the superalgebra of $\mathfrak{osp}(2|2)$.

The BRST invariant action is:
\begin{eqnarray}
\label{EQD1}
{\cal L}^{(3)} = {\cal L}^{(3)}_+ - {\cal L}^{(3)}_- \,,
\end{eqnarray}
where
\begin{eqnarray}
\label{EQE}
{\cal L}^{(3)}_+ &=& \frac12 \left( \omega_+^{\a\b} \wedge d \omega_{+, \a\b} -
\frac13 \omega_+^{\a\a'}\wedge \omega_{+,\a'\b'} \wedge \omega_+^{\b'\b}
\right) - 2\,i\, \epsilon_{\a\b}  \psi^{\a I} \nabla \psi^\b_I - A \wedge dA ~~~ ,\nonumber \\
\nabla \psi^\a_I &=& (\delta^\a_\b d + \frac12 \omega^\a_{~\b})\psi^\b_I -  \frac12 \epsilon_{IJ} A \wedge \psi^\a_J\,, \nonumber \\
{\cal L}^{(3)}_- &=& \frac12 \left( \omega_-^{\a\b} \wedge d \omega_{-,\a\b} -
\frac13 \epsilon_{\a\b} \omega_-^{\a\a'}\wedge \omega_{-, \a'\b'} \wedge \omega_-^{\b'\b}
\right) .
\end{eqnarray}
The first piece ${\cal L}^{(3)}_+$ is the Chern-Simons action related to the superalgebra
$\mathfrak{osp}(2|2)$,
while the second piece ${\cal L}^{(3)}_-$ is related to the bosonic algebra $\mathfrak{so}(1,2)$.

\subsection{A secondary BRST symmetry}

From \cite{Rozansky:1996bq}, trying to understand the origin of the world-volume
supersymmetry, we learn that there is a secondary BRST symmetry, that we denote by $\bar s$.
It is obtained by exchanging the role of the ghost field $\phi^\a_I$ with that of the anti-ghost
$\bar \phi^I_\a$, as follows:
\begin{eqnarray}
\label{EQCA_bar}
&&\bar s\, \omega^{\a\b}_- =0\,, ~~~~\nonumber \\
&&\bar s\,  \omega^{\a\b}_+ =- 2 \,i\,\bar\phi^{(\a}_I \psi^{\b)}_I  \,, ~~~~\nonumber \\
&&\bar s\,  A = - i\, \epsilon_{\alpha\beta}\,\epsilon_{IJ} \bar\phi^\a_{ I} \psi^\b_J\,, ~~~~\nonumber \\
&&\bar s\, \psi^\a_I =  {\cal D} \bar\phi^\a_{ I}
- \frac12 A \epsilon_{I J}  \bar\phi^\a_{ J} \equiv
\nabla  \bar\phi^\a_{ I} \,, \nonumber \\
&&\bar s\, \bar\phi_\a^I =0\,.
\end{eqnarray}
The Chern-Simons action and the fermionic terms are invariant under this BRST symmetry in the same way as
they are invariant under the BRST symmetry $s$. Just as in the latter case,  the nilpotency of the $\bar s$-BRST transformations is satisfied up to gauge transformations, with parameters $-i\,\mu^-_{(\a\b)}$ and $-i\,\mu^-$.
The analysis is performed along the same lines
as in (\ref{EQCD}). The indices in $\mu^-_{(\a\b)}$ appear in the lower position, but they can be raised
by the Cartan-Killing metric of the superalgebra: $\mu_{\a\b} = - \epsilon_{\a\gamma} \epsilon_{\b\delta} \mu^{\gamma\delta}$.

In addition, the $\bar s$ transformation of $\phi^\a_I$ is
\begin{eqnarray}
\label{EQCE_bar}
\bar s\, \phi_I^\a =-\bar \eta^{\a}_I\,, ~~~~~~
\bar s\, \bar \eta^{\a}_{ I} = -
\frac i2
(\mu_- \epsilon_{IJ}  \phi^{\a}_J +\mu_-^{(\a\b)}\epsilon_{\b\gamma}\phi^{\gamma}_I )\,.
\end{eqnarray}
Again, by computing the nilpotency of this new BRST differential $\bar s$, we see
that eq.s (\ref{EQCA_bar}) and (\ref{EQCE_bar}) close on gauge transformations (the supergauge transformations
induced by the supergroup ${\rm Osp}(2,2)$) with  parameters $-i\,\mu^-, -i\,\mu^-_{(\a\b)}$.  In order to check
the nilpotency, the conditions conjugated to those in (\ref{EQCCantiA}) are used.

The two BRST symmetries have to be compatibile. To this aim we need to check the
anticommutation relations between them. It is easy to show that we have
\begin{eqnarray}
\label{comm_SS}
\frac{1}{2}(s \, \bar s + \bar s s) =-\frac{i}{2} \mu_{3}^{\a\b} \delta_{(\a\b)}-i\,  \mu_{3} \delta\,,
\end{eqnarray}
where $\delta$ and $\delta_{(\a\b)}$ are the generators of the gauge symmetries $SO(2)$ and
$Sp(2)$ of the supergroup and $\mu_3^{(\a\b)}, \mu_3$ are the moment maps
related to K\"ahler structure $K$, introduced in (\ref{GFC}). This means that the anticommutation of the two BRST transformations yields a gauge transformation with parameters $-i\,\mu_3^{\a\b},\,-i \mu_3$, and therefore they anticommute
only when acting on gauge-invariant quantities.
\subsection{Vector BRST Symmetry}
Before entering into the detail of the  gauge fixing, which will be the subject of next section, let us clarify here where the world-volume supersymmetry on the gauge fixed theory comes from.
Even though a general discussion of that issue for all possible gauges  will deserve a longer work \cite{long_paper}, let us observe that,
in the BRST-gauge fixing of Chern-Simons theories,  supersymmetry emerges in a very interesting way. Indeed, the
action depends upon the world-volume metric only through the gauge-fixing term. The latter, as we shall discuss in the next section, is BRST exact and has  the general form $\int s\Psi$, where $\Psi$ is the so-called \emph{gauge-fixing fermion}, namely a fermionic function of the fields which encodes the gauge fixing.
Therefore,
the world-volume energy-momentum tensor satisfies the equation
\cite{Delduc:1990je,Vilar:1997fg,DelCima:1998ux,Gieres:2000pv}
\begin{eqnarray}
\label{ZANB}
\frac{\delta S}{\delta g_{\mu\nu}} \equiv T_{\mu\nu} = s \Gamma_{\mu\nu}
\end{eqnarray}
where $\Gamma_{\mu\nu}$ is the variation of the gauge-fixing fermion $\Psi$ with respect to the world-volume metric
$g_{\mu\nu}$.
It can be proven that the conservation of $\Gamma_{\mu\nu}$ (up to a suitable redefinition of $T_{\mu\nu}$)
follows from the following equation:
\begin{eqnarray}
\label{ZANC}
\partial^\mu \Gamma_{\mu\nu} =
\delta_\nu A_{\mu}^{\a\b} \frac{\delta S}{\delta A^{\a\b}_\mu} +
\delta_\nu A_{\mu, [IJ]} \frac{\delta S}{\delta A_{\mu, [IJ]}} +
\delta_\nu \psi_{\mu, I}^{\a} \frac{\delta S}{\delta \psi^{\a}_{\mu, I}} +
\delta_\nu \bar\eta^{I}_{\a} \frac{\delta S}{\delta \bar\eta_{\a}^{I}} +
\delta_\nu \phi_{I}^{\a} \frac{\delta S}{\delta \phi^{\a}_{I}} +
\delta_\nu \bar\phi^{I}_{\a} \frac{\delta S}{\delta \bar\phi_{\a}^{I}} \nonumber \\
\end{eqnarray}
which implies the existence of a rigid vector BRST-symmetry $\delta_\nu$. The form of the  field variations, whose explicit realization depends on the gauge-fixing considered,
can be read off from the various terms $\delta_\nu A_{\mu}^{\a\b}, \dots, \delta_\nu \bar\phi^{I}_{\a}$.

Moreover, a further vector BRST-symmetry transformation  $\bar\delta_\mu$ leaves invariant the gauge-fixed Chern-Simons lagrangian, as it can be checked for the Laundau gauge-fixing (\ref{GFA}), which will be discussed in Section 3.2; in fact, it can be rewritten in terms
of the $\bar s$ as follows $\int s \Psi = \int \bar s \bar \Psi$ where the ghost $\phi^\a_I$ and the anti-ghost
$\bar \phi_I^\a$ are exchanged, but this implies the existence of a further vector BRST-symmetry, $\bar\delta_\mu$. We shall denote by $s_\mu,\,\bar{s}_\mu$ the abstract generators of $\delta_\mu,\,\bar{\delta}_\mu$, respectively.

All the symmetries
are recombined into an $\mathcal{N}=4$ supersymmetry formulation \cite{DelCima:1998ux}, to be compared with the twisted version of the $\mathcal{N}=4$ supersymmetry of the dual model. Indeed, the scalar supersymmetries represented by the BRST operators $s$ and $\bar s$ and the vector supersymmetries $s_\nu$ and $\bar s_\nu$ are naturally combined in the ${\cal N}=4$ supersymmetry mentioned
earlier.  We shall elaborate further on this at the end of subect. \ref{countingdof}.

\section{Gauge-Fixing Choices}

\subsection{Counting of D.O.F.'s}\label{countingdof}

Before discussing the gauge-fixing, it is convenient to count the off-shell
degrees of freedom. This will clarify the correspondence, outlined in the Introduction, between the Achucarro-Townsend model \cite{Achucarro:1987vz} and an  $\mathcal{N}=4$ world-volume supersymmetric Chern-Simons theory coupled to matter, analogous to the model discussed in \cite{Gaiotto:2008sd}. In the counting, negative d.o.f.'s mean gauge symmetries.

We have the gauge fields $\omega^{(\a\b)}_{\pm,\mu}$, each of them having  $ (3 \times 3 -3)$
d.o.f.'s, the fermionic gauge fields $\psi^\a_{I,\mu}$, with $(2 \times 2 \times 3 - 2 \times 2)$ and the $\mathrm{SO}(2)$ gauge field $A_\mu$ with $(3 -1)$ d.o.f.'s.
We note that, in the supersymmetric sector,
 the bosonic d.o.f.'s ($6+2$) match the fermionic ones.

In addition, we note that  $\psi^\a_{I,\mu}$ is a 1-form carrying one index in the $\mathrm{Sp}(2)$ fundamental representation and one index in the $\mathrm{SO}(2)$ R-symmetry vector representation of the bosonic symmetry in the gauge supergroup. Therefore, at first sight it does not have the features of
a  Rarita-Schwinger field on the world-volume. The interpretation as a gravitino
follows from the identification of the subgroup $\mathrm{Sp}(2) \sim \mathrm{SO}(1,2)$ of the gauge supergroup with
the world-volume $\mathrm{SO}(1,2)'$ Lorentz group.

In the AVZ model \cite{Alvarez:2011gd}, the authors introduce a spinor $0$-form $\chi^{\b'}_I$ as in eq. (\ref{zanelliwv}),
where the index $\beta'$ is a truly spinorial index on the world-volume, and $\gamma^{\a\beta'}_i$
are the Dirac matrices which intertwine between the gauge $\mathrm{SO}(1,2)$ group
and the world-volume Lorentz  $\mathrm{SO}(1,2)'_L$ group, as discussed in sect. \ref{girella}. The matrix $e^i_\mu$ is the 3\,d dreibein, associated with the adjoint representation of the diagonal subgroup $Sp(2)'_D\subset SO(2,2)'$ of the world-volume isometry group.

However, it would be desiderable to derive (\ref{zanelliwv}) in terms of a gauge symmetry of the model (which actually reduces the
off-shell degrees of freedom from $8$ down to $4$). To this end, let us observe that
the felds $\psi^\a_I$ are fermionic 1-forms, therefore their components $\psi^\a_{I, \mu}$ are
fermionic d.o.f.'s.
In the ghost sector we introduce the Nakanishi-Lautrup fields $\bar\eta^I_\a$
needed for the gauge-fixing of the fermionic gauge symmetry. The total amount of fermionic degrees of freedom is
then $ 2 \times 2 \times 3$ for  $\psi^\a_{I, \mu}$ \footnote{Indeed, in the BRST-invariant gauge-fixed action, all d.o.f.'s are propagating.}  and $2 \times 2$ for $\bar\eta^I_\a$.

All the off-shell fermionic d.o.f.'s in our gauge-fixed 
model can be arranged into a single spinor field
of the form ${\Lambda}^\a_{I, \a'\b'}$ given by
\begin{eqnarray}
\label{EQI}
\Lambda^{\a}_{I \a' \b'} = i \gamma^{\mu}_{\a' \b'} \psi^\a_{\mu, I} - \frac12
\epsilon_{\a' \b'}  \,  \bar\eta^\a_{I}\,,
\end{eqnarray}
where $\a'$ and $\b'$ both refer to the diagonal world-volume symmetry $Sp(2)'_D\subset SO(2,2)'$. This is the analogue in our setting of the topological twist which was shown in \cite{Kapustin:2009cd} to relate the Gaiotto-Witten model \cite{Gaiotto:2008sd} with a gauged version of the Rozansky-Witten one \cite{Rozansky:1996bq}.
On the other hand, we can perform a different twisting in order to make contact with the AVZ model and to identify its fermionic degrees of freedom in the present context. This is achieved by decomposing $\Lambda^{\a}_{I \a' \b'}$ with respect to the diagonal subgroup of the target space
$\mathrm{Sp}(2)$ (which the index $\alpha$ refers to) and of the world-volume $Sp(2)'_D$,
the fermionic d.o.f.'s
$\Lambda^\a_{I, \a'\b'}$, so as to obtain
\begin{eqnarray}
\label{EQI_A}
\Lambda^{\a}_{I \a' \b'} = i (\gamma^{\mu})^\a_{~\a'} \chi_{\mu \b', I} - 2 \delta^\a_{~\a'} \chi_{I\b'}\,.
\end{eqnarray}
On the right hand side the field $\chi_{\mu \b', I}$ is identified with the world-volume gravitino, while $\chi_{I\b'}$ contains the AVZ fermionic degrees of freedom.

The full world-volume isometry $SO(2,2)'$ can be made manifest by promoting $\Lambda^{\a}_{I \a' \b'}$ to an object in the $(1/2,1/2,1/2)$ of $Sp(2)\times Sp(2)_+' \times Sp(2)_-'$:
\begin{equation}
\Lambda^{\a}_{I \a' \b'}\longrightarrow \Lambda^{\a}_{I \a' \dot\b'} \label{full}
\end{equation}
where we recall that $\a'$ and $\dot\b'$ refer to the groups $Sp(2)'_+$ and $Sp(2)'_-$, respectively.

The on-shell analysis is difficult since there are no local degrees of freedom.
In addition, in the previous sections we introduced the ghost fields $\phi^\a_I, \bar\phi^I_\a$, that are scalar bosonic degrees of freedom, which are interpreted as the scalar superpartners of $\Lambda^{\a}_{I \a' \b'}$.
As discussed earlier, they naturally parametrise a hyper-K\"ahler manifold.

We can then group the d.o.f.'s as follows: the gauge fields
$\omega^{(\a\b)}_\pm$ and $A$ for the gauge group $\mathrm{Sp}(2)\times \mathrm{Sp}(2) \times \mathrm{SO}(2)$ are described by a Chern-Simons gauge theory, while the other fields $\phi, \bar\phi$ and $\Lambda$ build up an $\mathcal{N}=4$ hypermultiplet charged with respect
to the gauge fields. The choice of a suitable potential allows to recast the model into a $\mathcal{N}=4$ super-Chern-Simons theory. The corresponding ${\cal N}=4$ supersymmetry parameter has the following structure:
$\epsilon^{\alpha'\dot\alpha' A}$, where the $SU(2)$ group acts on the index $A$ \cite{long_paper}. As discussed in Sect. 2, this emerging supersymmetry is related, via the topological twist discussed above, to the scalar and vector BRST quantum symmetries of the model: $s,\,\bar{s},\,s_\mu,\,\bar{s}_\mu$.
This can be seen by decomposing the corresponding supersymmetry parameters
$\epsilon^{\alpha' \dot\alpha'\ A}$ with respect to $Sp(2)'_D \times SU(2)$:
\begin{equation}\epsilon^{\alpha' \dot\alpha' A} = i (\gamma^{i})^{\alpha'\dot\alpha'} \epsilon^A_i +
\epsilon^{\alpha'\dot\alpha'} \epsilon^A\,,\end{equation}
 where $\epsilon^{A=1}, \epsilon^{A=2}$ correspond to $s$ and
$\bar s$ BRST symmetries and $\epsilon^{A=1}_i, \epsilon^{A=2}_i$ correspond to vector supersymmetries
$s_\mu$ and $\bar s_\mu$.

\subsection{Landau Gauge-Fixing and $\mathcal{N}=4$ supersymmetry}

The gauge-fixed Lagrangian (only for the fermionic gauge symmetry) has the general form
\begin{eqnarray}
\label{EQF}
{\cal L} = {\cal L}^{(3)} + {\cal L}^{g.f.}\,, ~~~~~
{\cal L}^{g.f.}=  s
\Big( \bar \phi^I_\a {\mathcal F}_I^\a(\omega_+, \psi_I^\a, A, \phi_I^\a) \Big)\,, ~~~~~
S = \int_{{\cal M}_3} {\cal L}\,,
\end{eqnarray}
where $\Psi = \bar \phi^I_\a  {\mathcal F}_I^a(\omega_+, \psi_I^\a, A, \phi_I^\a)$ is known as the {\it gauge-fixing fermion}
which encodes  the gauge-fixing and depends
upon the gauge fields $(\omega_+, \psi_I^\a, A)$ and possibly also on the ghost field $\phi^\a_I$.

Let us discuss the gauge-fixing in detail.
As already mentioned in the introduction, we do not introduce here the ghosts for the bosonic part of the gauge symmetry, since we
are only interested in the gauge-fixing of the odd part of the superalgebra. The gauge-fixing
of the even part of the gauge algebra can be done according to the standard procedure.

Let us start with a gauge-fixing, known also as Landau gauge fxing, of the following form (up to total derivatives)
\begin{eqnarray}
\label{GFA}
S_A = 2\int s \Big( \bar\phi^I_\a  \nabla  \star  \psi^\a_I \Big) &=&-2
\int \Big( \nabla \phi^\a_I \wedge \star \nabla \bar\phi^I_\a +
\psi^\a_I \wedge \star \nabla \bar \eta^I_\a \Big)\nonumber \\
&&+ 2\int \Big( - i \,  \phi^{J (\a} \psi^{\b)}_J \bar\phi^I_{\a} \star \psi_{\b I} - \frac12 \epsilon^{KL} \phi^\gamma_K \psi^\delta_L \epsilon_{\gamma \delta } \bar\phi^I_\a \epsilon_I^{~J} \star \psi^\a_J \Big) \,.
\end{eqnarray}
The Hodge dual $\star$ is needed to write the gauge fixed action on any three dimensional manifold, given
a metric on it. In particular, we have: $\psi^\a_I \wedge \star \psi^\b_J =
\psi^\a_{\mu, I} g^{\mu\nu} \psi^\b_{\nu, J} V^{(3)}$,
where $g^{\mu\nu}$ is the inverse metric on the world-volume ${\cal M}^{(3)}$,
whose volume form is $V^{(3)} = \star 1$. Notice that a world-volume metric is needed only in the gauge-fixing term and it is not present in the
gauge-invariant action ${\mathcal L}^{(3)}$.

The first term in (\ref{GFA}) is an usual kinetic term for complex bosonic fields $\phi^\a_I, \bar\phi^I_\a$.
The second term contains a
first-order linear differential operator on $\bar \eta^I_\a$ which, together with the gravitino field
equation from eq. (\ref{EQE}), leads to an invertible wave operator. The last term
produces an interaction between $\psi$ and the ghost fields
$\phi$ and $\bar \phi$. Notice that the ghost fields in the present case are bosonic
commuting fields, therefore the kinetic term leads to a positive definite metric for
the Hilbert space.

Now, we check that the gauge-fixing fixes the fermionic gauge symmetry and we study the wave operator in the
fermionic sector. In order to simplify the discussion,
we neglect for the time being the interactions among ghost fields $(\phi^\a_I, \bar \phi^I_\a)$.

The free equations for $\psi^\a_I$ and  for
$\bar\eta^I_\a$ read
\begin{eqnarray}
\label{EQA}
2 \epsilon_{\a\b}  \nabla \psi^\b_I + \star \nabla \bar\eta_{I\a} =0\,, ~~~~~
\nabla \star \psi^\a_I =0\,.
\end{eqnarray}
We re-write these equations in components as follows
\begin{eqnarray}
\label{EQB}
 \epsilon^{\mu\nu\rho}   \epsilon_{\a\b} \nabla_\nu \psi_{\rho, I}^\a + \nabla^\mu \bar\eta_{I\a} =0\,, ~~~~~~
\nabla^\mu \psi_{\mu, I}^\a =0\,.
\end{eqnarray}
The second equation is the usual Landau-Lorentz gauge-fixing for the gravitino.

By standard manipulations, using the Clifford algebra on the world-volume and the gauge-fixing condition,  
we have
\begin{eqnarray}
\label{EQD}
2   (\gamma_\mu)_{\a\b} \nabla^\mu (\gamma^\nu \psi_{\nu, I})^\b + (\gamma_\mu)_{\a}^{~\b}
\nabla^\mu \bar\eta^I_\beta =
\not\!\nabla \Big( 2 \epsilon_{\a\b} (\not\!\!\psi^\b_{I}) + \bar\eta^I_\alpha\Big) =
0
\end{eqnarray}
form which we get the following combination
\begin{eqnarray}
\label{ZANA}
\bar\eta^I_\alpha = - 2  (\not\!\!\psi_I)_\a + \sigma^I_\a
\end{eqnarray}
where $\sigma^I_\a$ is a solution of the massless Dirac equation $\not\!\!\partial \sigma^I_\a =0$.
%
However, it is possible to further modify the gauge-fixing in order to introduce a mass term for the Dirac field.


\subsection{Feynman Gauge Fixing and Mass Deformations}

The Feynman gauge-fixing also requires the introduction of the Nakanishi-Lautrup auxiliary field
$\bar\eta^\a$. It can be added to the gauge-fixing as follows
\begin{eqnarray}
\label{GFB}
S_B &=& 2\int s \Big( \beta \bar\eta^I_\a  \epsilon^{\a\b} \bar \phi_{\b,I} \Big) V^{(3)} \nonumber \\
&=&2
\int \beta  \Big( \bar \phi^I_\a (\mu_{+, IJ}(\phi) \epsilon^{\a\b} + \mu_+^{\a\b}(\phi) \delta_{IJ}) \bar \phi^J_
\b + \bar\eta^I_\a \epsilon^{\a\b} \bar \eta_{\b,I} \Big) V^{(3)}\nonumber \\
&=&
2\int \beta \Big( \mu_{3,\a}^{~~~\b} \mu_{3, \gamma}^{~~~\delta} \epsilon_{\a\gamma} \epsilon^{\b\delta} +
\mu_{3,K}^{~~~I} \mu_{3,I}^{~~~K} + \bar\eta^I_\a  \epsilon^{\a\b} \bar \eta_{\b,I} \Big) V^{(3)}\,.
\end{eqnarray}
Here the first two terms in the last line are potential terms for the bosonic ghost fields $(\phi, \bar\phi)$,
while the third one is a mass term for $\bar \eta$. We used the definition of $\mu_3$ to simplify the
expression and $\beta$ is a gauge-fixing parameter.

The form of the potential generated by the present gauge-fixing is similar to the mass deformations
of the Gaiotto-Witten model discussed in \cite{Hosomichi:2008jb,Koh:2009um}. As mentioned in the Introduction, the model we are considering bears resemblance with the ABJM model \cite{Aharony:2008ug}
in that it is described by the difference of two Chern-Simons.
However, as is apparent from (\ref{GFB}), the BRST variation
of this gauge-fixing term cannot generate higher powers of $\phi$ and $\bar\phi$ which could reproduce
the sextic scalar potential of ABJM models.

The gauge-fixing produces a $\bar\eta^2$ term,
which serves as the Nakanishi-Lautrup term, modifying the quadratic part of the $\psi-\bar \eta$ system,
whose equations of motion were discussed in the previous section.

\subsection{Non-linear Feynman gauge-fixing and Gaiotto-Witten model}

One can add further interaction terms by introducing a more general  gauge-fixing piece of the form
\begin{eqnarray}
\label{GFD}
S_B' &=& \int s \Big(  \bar\eta^I_\a V^{\a\b}_{IJ}(\phi) \bar \phi^J_\b \Big) V^{(3)} \nonumber \\
&=& \int  \Big( \bar \phi^K_\a \bar \phi^J_\b (\mu_{+, K}^{~~~I} V_{IJ}^{\a\b} +
\mu_{+, \gamma}^{~~~\a} V^{\gamma \b}_{IJ}) + \bar \eta^I_\a V^{\a\b}_{IJ} \bar\eta^J_\b
\Big) V^{(3)}\,,
\end{eqnarray}
where $V^{\a\b}_{IJ}$ is a generic tensor built out of the $\phi$'s. As can be noticed, this additional term
modifies the action by non-quadratic terms and it is responsible of the generation of the scalar potential.
As discussed in \cite{Kapustin:2009cd}, a suitable choice of $V^{\a\b}_{IJ}(\phi)$ (compatible with the bosonic gauge
invariance of the model) leads to the scalar potential of the dual theory and the coupling with
the fermions.

\subsection{Unconventional Gauge Fixing and AVZ Ansatz}

Let us now add a new term to the equations of motion and recompute the wave operator. In particular,
we will derive the AVZ Ansatz for a massive Dirac spinor by a suitable choice of the gauge-fixing parameters.
The corresponding quadratic part of the Lagrangian is
\begin{eqnarray}
\label{EQAIB}
{\cal L} =2 \epsilon_{\a\b} \epsilon^{\mu\nu\rho}
\psi^\a_{\mu, I} \nabla_\nu \psi^\b_{\rho, I} + \bar\eta^I_\a (\nabla^\mu \psi^\a_{\mu, I} - 2 i \beta
(\gamma^i)^\a{}_{\b} \psi^\b_{i, I}) + 2\beta  \epsilon^{\a\b} \, \bar\eta^I_\a \bar\eta_{\b, I}
\end{eqnarray}
where the last term is the usual quadratic term in the Nakanishi-Lautrup field $\bar\eta^I_\a$.\footnote{In the case of usual gauge symmetry the associated BRST symmetry is $s\, A = \nabla c, s\, c = - \frac12 [c,c], s\, \bar c = b,  s\, b =0$ where $b$ is the Nakanishi-Lautrup field. Then, the gauge-fixing Lagrangian is
$s\,  {\rm Tr}( \bar c \, d\star A + \frac12 \xi \bar c b) = {\rm Tr} (b  \, d\star A + \frac12 \xi b^2) - {\rm Tr}
(\bar c \, d\star \nabla c)$. The equations of motion for $b$ is $b = - \frac{1}{\xi} d\star A$ which implies that
gauge-fixing for $A$.}

The new equations of motion are
%
\begin{eqnarray}
\label{EQAII}
4
\epsilon^{\mu\nu\rho} \nabla_\nu \psi_{\rho, I}^\a - 2 \nabla^\mu \bar\eta^\a_I +
4 i \b \gamma^{\mu, \a}{}_{\b} \bar\eta_I^\beta =0\,, \nonumber \\
\nabla^\mu \psi_{\mu, I}^\a - 2 i \b (\gamma^\mu \psi_{\mu,I})^\alpha
+ 2 \b \epsilon^{\a\b}  \bar\eta^I_\b =0\,.
 \end{eqnarray}
Manipulating those equations as above,
we finally find the massive Dirac equation for the combination
$\chi_I = -\frac 14 (i\,\not\!\!\psi_I + \frac 12  \bar\eta_I)$ which should be identified with the AVZ Dirac spinor.

Finally, we can write down the complete {\it unconventional} gauge-fixing of the CS action with this additional piece as follows
\begin{eqnarray}
\label{GFaddA}
S_A &=& \int s \Big( \psi^\a_I \wedge \star \nabla \bar\phi^I_\a  +
\psi^\a_I \wedge \star e^a \gamma_{a, \alpha}^\beta \bar\phi^I_\b \Big)
\nonumber \\
&=&
\int \Big[\nabla \phi^\a_I \wedge \star \nabla \bar\phi^I_\a +
\psi^\a_I \wedge \star \nabla \bar \eta^I_\a
+
\psi^\a_I \wedge \star \Big( \epsilon_{\a\b} \psi^{(\b}_K \eta^{KL} \phi^{\gamma)}_L \bar \phi^I_\gamma
+ \eta^{IJ} \psi^\gamma_{[J} \epsilon_{\gamma\beta} \phi^\b_{K]} \bar\phi^K_\alpha \Big)
\nonumber \\
&& +   \nabla \phi^\a_I \wedge \star e^i \gamma_{i, \alpha}^\beta \bar\phi^I_\b  +
\psi^\a_I \wedge  \star e^i \gamma_{i, \alpha}^\beta  \bar \eta^I_\b \Big].
\end{eqnarray}
No additional term is produced by the BRST variation of the second term of the gauge-fixing.
In addition, no BRST variation is required for the 3d vielbein $e^i$. The new term leading to the unconventional
gauge-fixing and to the identification of our fermionic fields with the AVZ Ansatz modifies the kinetic
term of the bosonic ghost fields (viewed as  coordinates of a hyper-K\"ahler manifold) by a vorticity term involving a first order differential.

\subsection{$s \bar s$-Gauge Fixing}

Finally, let us come to the last example of gauge-fixing that can be constructed. Given the
fact that there are two BRST symmetries, instead of writing the gauge-fixing
as the $s$-variation of $\Psi$, one can construct the gauge-fixing as
follows
\begin{eqnarray}
\label{ss-gaA}
{\cal L}^{g.f.} = s\bar s \Sigma
\end{eqnarray}
where $\Sigma$ has no ghost charge, it is Lorentz invariant and it must
also be gauge invariant under the bosonic subgroup of the supergroup ${\rm OSp}(2|2)$.
Since $s$ and $\bar s$ are nilpotent, the total action $S$ will be $s$ and $\bar s$ invariant.
Acting with $s$, it is trivially zero, while to check the $\bar s$-invariance, we have to anticommute $\bar s$ with $s$ and that can be
done at the price of a gauge transformation in the subgroup. However,  being $\Sigma$ gauge invariant,
this gauge transformation is ineffective and nilpotency of $\bar s$ does the rest.

The Landau gauge-fixing plus non-linear terms (\ref{GFA})-(\ref{GFD}) can be written as follows
\begin{eqnarray}
\label{new_gauge}
{\cal L}^{g.f.} = s\bar s \Big( \psi^\a_I \epsilon_{\a\b} \eta^{IJ}\wedge \star \psi^\b_J +
\bar \eta_\a^I \epsilon^{\a\b} \bar \eta_\b^I \star 1\Big).
\end{eqnarray}
Notice that those terms are gauge invariant under the bosonic gauge symmetries of the
supergroup ${\rm SO}(2)$ and ${\rm Sp}(2)$ and, therefore, as pointed out earlier,
the order of $s$ and $\bar s$
in front of the parenthesis is irrelevant. We have made explicit the contraction
of the indices.

Using the present framework, we can rewrite the unconventional gauge-fixing in a similar
way by observing that since the group ${\rm Sp}(2)$ is the spin group associated with ${\rm SO}(1,2)$
(the Lorentz group of the world-volume), we can use the intertwiners $\gamma^{\a\b}_i$
(or even better $\Gamma_{i\b}^{~~\a} = i \epsilon_{\a\gamma} \gamma^{\gamma\b}_i$)
to extend the above gauge-fixing term as follows
\begin{eqnarray}
\label{new_gauge2}
{\cal L}^{g.f.} = s\bar s \Big( \psi^\a_I \epsilon_{\a\b} \wedge \star \psi^\b_I +
\bar \eta_\a^I \epsilon^{\a\b} \bar \eta_{\b |I} \star 1+
\beta \, \psi^\a_I \wedge \star e^i \Gamma_{i\a}^{~~\b}  \bar \eta^I_\b \Big)
\end{eqnarray}
where $e^i$ is the wordvolume vielbein and $\star e^i$ is the world-volume Hodge dual and
$\beta$ is the gauge-fixing parameter discussed in the previous section.
This beautiful structure contains all possibile quadratic terms written in terms of only the
fermionic fields of the theory, namely $\psi^\a_I$ and $\bar \eta^I_\a$. They are gauge invariant
under the bosonic symmetries of the supergroup together with the Lorentz transformations on the
flat index of $e^i$.

To conclude, we observe that we can add a further term
in order to build a quadratic term, of the form
\begin{eqnarray}
\label{SM_QA}
s\bar s \Big( A^{(\a\b)} \gamma_{i \a\b} \wedge \star e^i \Big)
\end{eqnarray}
which carries zero ghost number and is a 3-form. It is manifestly Lorentz invariant and its variation
gives
\begin{eqnarray}
\label{SM_QB}
s \Big( \psi^{\a} \wedge \bar\phi^\b (\gamma_i)_{\a\b} \star e^i \Big) =
\nabla \phi^\a \wedge \bar\phi^\b (\gamma_i)_{\a\b} \star e^i -  \psi^{\a} \wedge \bar\eta^\b  (\gamma_i)_{\a\b} \star e^i \,.
\end{eqnarray}
The first term is a viscosity term modifying the quadratic part of the action for the scalar fields. The second term
is an off-diagonal mass term for the fermions.

Therefore, also in the present formulation, there are at least two supercharges in the game and part of the supersymmetry
is preserved. We leave further consideration in a forthcoming more detailed work \cite{long_paper}.

\section{Conclusions and Outlook}

In the present work,
we have considered different ways of quantizing Chern-Simons supergravity revealing
important features of the same theory.
This is achieved by applying the analysis of Kapustin and Saulina \cite{Kapustin:2009cd} to this particular Chern-Simons theory describing pure $D=3$ supergravity on an ${\rm AdS}_3$ background.\par The novel feature which characterizes our setting is indeed the presence of a curved background with negative cosmological constant. In this framework both the fermionic degrees of freedom of the AVZ model \cite{Alvarez:2011gd} and the topological twist of \cite{Kapustin:2009cd} find a natural interpretation, the latter being related to the choice of the Lorentz ${\rm Sp}(2)$ world volume spin connection within the ${\rm SO}(2,2)$ isometry group of space-time. \par
Our analysis led us to consider different choices of the gauge fixing of the fermionic symmetries.
The resulting models exhibit some amount of supersymmetry, which emerges from the BRST quantization.
The ultimate goal, which will be tackled in a forthcoming
work \cite{long_paper}, is the study of the supersymmetry associated with the different gauge-fixing choices. In this respect let us summarize the results:
\begin{enumerate}
\item Landau Gauge Fixing: this is achieved by setting to zero the Nakanishi-Lautrup terms of the gauge-fixing (\ref{GFB})
and non-linear gauge fixing terms (\ref{GFD}). The model has a manifest $\mathcal{N}=4$ supersymmetry as a result of a combination
of BRST symmetries ($s$ and $\bar s$) and of the vector supersymmetries.
\item Kapustin-Saulina Gauge Fixing: adding the non-linear terms (\ref{GFD}). Those terms are needed to reproduce the
scalar potential of the dual theory. In the dual picture, the theory still displays an ${\cal N}=4$
supersymmetry, though realised in a non-linear way.
That supersymmetry is a manifestation of the BRST symmetries of the original Chern-Simons supergravity and of the vector supersymmetris, which however are not evident in the twisted version due to the non-linearities. This leads to an open
problem: to show that the CS supergravity with this particular gauge-fixing admits a formulation with a manifest non-linear vector supersymmetry.
\item Gauge Fixing with Mass Terms: in order to reproduce the deformations discussed in \cite{Hosomichi:2008jd,Hosomichi:2008jb,Koh:2009um} one has to change the gauge-fixing adding a new Nakanishi-Lautrup term. Its BRST variation leads to the expected terms.
\item Unconventional Gauge Fixing: As shown in the text, the choice of a particular gauge-fixing with a vorticity term, upon identification of the world-volume Lorentz group with the $SO(1,2)$ target gauge symmetry, leads to the AVZ Ansatz and to
the interpretation of the gravitino in terms of the graphene fermion. It is to be explored how much of the supersymmetry survives in the
present context.
\item $s\bar s$ Gauge-Fixing: choosing this form of the gauge-fixing there are less deformations allowed in the Lagrangian; nonetheless both the BRST and
the secondary BRST symmetry are present. In addition, the vorticity term discussed in the previous item is introduced by suitable
terms combining the gauge field and the world-volume dreibein. It would be very interesting to explore the consequences of this choice.

\end{enumerate}
The present analysis paves the way to further investigation in different directions: the detailed study of the supersymmetric properties of the quantum dual models originating from different gauge-fixings; the generalization of the present analysis to generic $N$-extended  $AdS_3$ supergravities; the investigation of the holographic $D=4$ duals to these models. As for the latter objective, a natural candidate would be the maximal $D=4$ supergravity realized on the $\mathcal{N}=4$ $AdS_4$ vacuum \cite{Gallerati:2014xra,Inverso:2016eet}, which describes a class of Type IIB Janus solutions \cite{DHoker:2007zhm,DHoker:2007hhe}.

As a concluding remark: we have shown that the $N$-extended D=3 supergravity can be reformulated
in terms of worldvolume supersymmetric models coupled to matter. This might be useful for
computation of interesting observables in both sides of this duality. We leave this to further investigations.

\section*{Acknowledgements}
We thank L. Castellani, R. D'Auria, P. Fr\'e, R. Olea, S. Penati, D. Seminara and J. Zanelli
for useful discussions on several issues of the present
work.




\end{document}